\documentclass[11pt]{article}
\usepackage{moriond,epsfig}

\def\be{\begin{equation}}
\def\ee{\end{equation}}
\def\bea{\begin{eqnarray}}
\def\eea{\end{eqnarray}}

\begin{document}
\vspace*{4cm}
\title{Propagation of extragalactic ultra-high energy cosmic-ray nuclei : implications for the observed spectrum and composition}

\author{Denis Allard}

\address{Laboratoire Astroparticule et Cosmologie (APC), CNRS/Paris VII \\
10 rue Alice Domon et L\'eonie Duquet, 75205 Paris Cedex 13, France.}

\maketitle\abstracts{In this paper, we study the propagation of cosmic-ray nuclei and protons. We emphasize the influence of the source composition on the expected spectrum and composition on earth as well as on the phenomenology of the transition from Galactic to extragalactic cosmic-rays and the interpretation of the ankle. We point out that the different source composition models cannot be distinguished on the sole basis of the cosmic-ray spectrum but that the the energy evolution of $\langle X_{\max} \rangle$ should remove this degeneracy. Finally, we compare the predictions for the different source composition models with the available data and discuss the implications of our results.}

\section{Introduction}
The cosmic-ray (CR) spectrum has been measured for half a century over 12 order of magnitude in energy and 32 in Flux. This spectrum shows an extraordinary regularity as it can almost be fitted by a single power law between $\sim 10^{10}$ and $10^{20}$ eV. One can however notice that this regularity is broken at least in two places : the so-called  "knee" (a steepening around 3-5$\times 10^{15}$ eV \cite{Kascade}) and the "ankle" (a hardening around 3-5$\times 10^{18}$ eV). These features are of considerable interest for astrophysics. At the highest energies a suppression of the flux seems to be observed independently by two experiments (HiRes \cite{HiResGZK} and Auger \cite{AugerSpectrum}) above $4-5\times 10^{19}$ eV probably corresponding to the expected GZK cut-off \cite{G66,ZK66}.
However, at high energy, composition analyses are difficult to perform due to the large hadronic model dependance of air showers properties.  As a result, a clear understanding of the highest energy cosmic-ray spectrum and its composition cannot yet be obtained. In particular, the question of the transition from Galactic (GCR) to extragalactic cosmic-rays (EGCR)  is under intense debates.  Moreover, a clear identification of the UHECR sources, crucial to constrain the acceleration mechanisms at play is not possible yet.  In this context, propagation studies of extragalactic ultra-high energy cosmic-ray (UHECR) nuclei can be very useful to identify spectral and composition signatures of different source composition models that can be compared to the available data.

In this proceeding, we study the propagation of cosmic-ray protons and nuclei. After presenting the relevant photon backgrounds and interaction processes relevant at the highest energies, we calculate propagated spectra and the expected energy evolution of $\langle X_{\max} \rangle$ under different assumptions for the source composition and cosmological evolution of the source luminosities.  We finally compare the model predictions with recent cosmic-ray data and discuss our results.




\section{Interactions of protons and nuclei with photon backgrounds}
Protons and atomic nuclei are usually considered as the most likely primary particles to explain the UHE cosmic-ray fluxes observed on Earth. Since the pioneer study of Puget, Stecker and Bredekamp \cite{PSB}, many authors have considered the propagation of UHE nuclei\cite{Rachen,Epele+98,Stecker99,Anchordoqui1998,Anchordoqui1999,Anchordoqui2001,Anchordoqui2007,Bertone+02b,Sigl2004,Sigl05,Gunter,Hooper2007,Hooper2008,Arisaka2007,Aloi2008,Berz2008,Med2008}. 

In the following sections we consider the interaction of protons and nuclei with the CMB and the infra-red, optical and ultraviolet backgrounds (hereafter we group these three backgrounds under IR/Opt/UV for short). To model the IR/Opt/UV backgrounds and their cosmological evolution, we use the latest estimate of \cite{MS05} which is based on  the earlier work of the authors updated with  recent data on history of the star formation rate and the evolution of galaxy luminosity functions.
We use IR/Opt/UV calculated at 26 different redshifts ($\Delta z=0.2$) between 0 and 5. 
  
Protons and nuclei propagating in the extragalactic medium interact with CMB and  IR/Opt/UV background  photons. These interactions produce features in the propagated UHECR spectrum such as the GZK cutoff \cite{G66,ZK66} and their decay products generate the cosmogenic neutrino flux \cite{BereOriginal}. 
In the case of protons the energy losses are dominated at low energy by adiabatic losses. Interactions with CMB photons become relevant at $\sim 10^{18}$ eV (at z=0) through the pair production process, dominant up to $\sim 7\times 10^{19}$ eV where the pion production, responsible for the GZK suppression, takes over. Interaction of protons with IR/Opt/UV photons are subdominant on the whole energy range.

The interactions experienced by nuclei with photon backgrounds are different from the proton case. Pair production (for which we use the mass and charge scaling given in \cite{Rachen}) and adiabatic losses result in a decrease of the Lorentz factor of the UHE nucleus, whereas  photodisintegration (also called photoerosion) processes lead to the ejection of one or several nucleons from the nucleus. Different photoerosion processes become dominant in the total interaction cross section at different energies \cite{PSB}.  The lowest energy disintegration process is the Giant Dipole Resonance (GDR) which results in  the emission of one or two nucleons and  $\alpha$ particles. The GDR process is the most relevant as it has the highest cross section with  thresholds  between 10 and 20  MeV for all nuclei. For nuclei with mass $A \geq 9$, we use  the theoretically calculated GDR cross sections presented in \cite{Khan04}, which take into account all the individual reaction channels and are in better agreement with data than previous treatments.  For nuclei with $A < 9$, we use the phenomenological fits to the data provided by \cite{Rachen}.
Around 30 MeV in the nucleus rest frame and up to the photopion production threshold, the quasi-deuteron (QD) process becomes comparable to the GDR and dominates the total cross section at higher energies.   The photopion production (or baryonic resonances (BR)) of nuclei becomes relevant  above 150 MeV in the nuclei rest frame  (e.g.,  $\sim5\times 10^{21}$ eV in the lab frame for iron nuclei interacting with the CMB), and we use  the parametrization given in \cite{Rachen} where the cross section in this energy range is proportional to the mass of the nucleus (nuclear shadowing effects are expected to break this scaling above 1 GeV). The reference for this scaling is the deuteron photoabsorption cross section which is known in great detail.

\begin{figure*}[ht]
\hfill\includegraphics[width=0.3\textwidth]{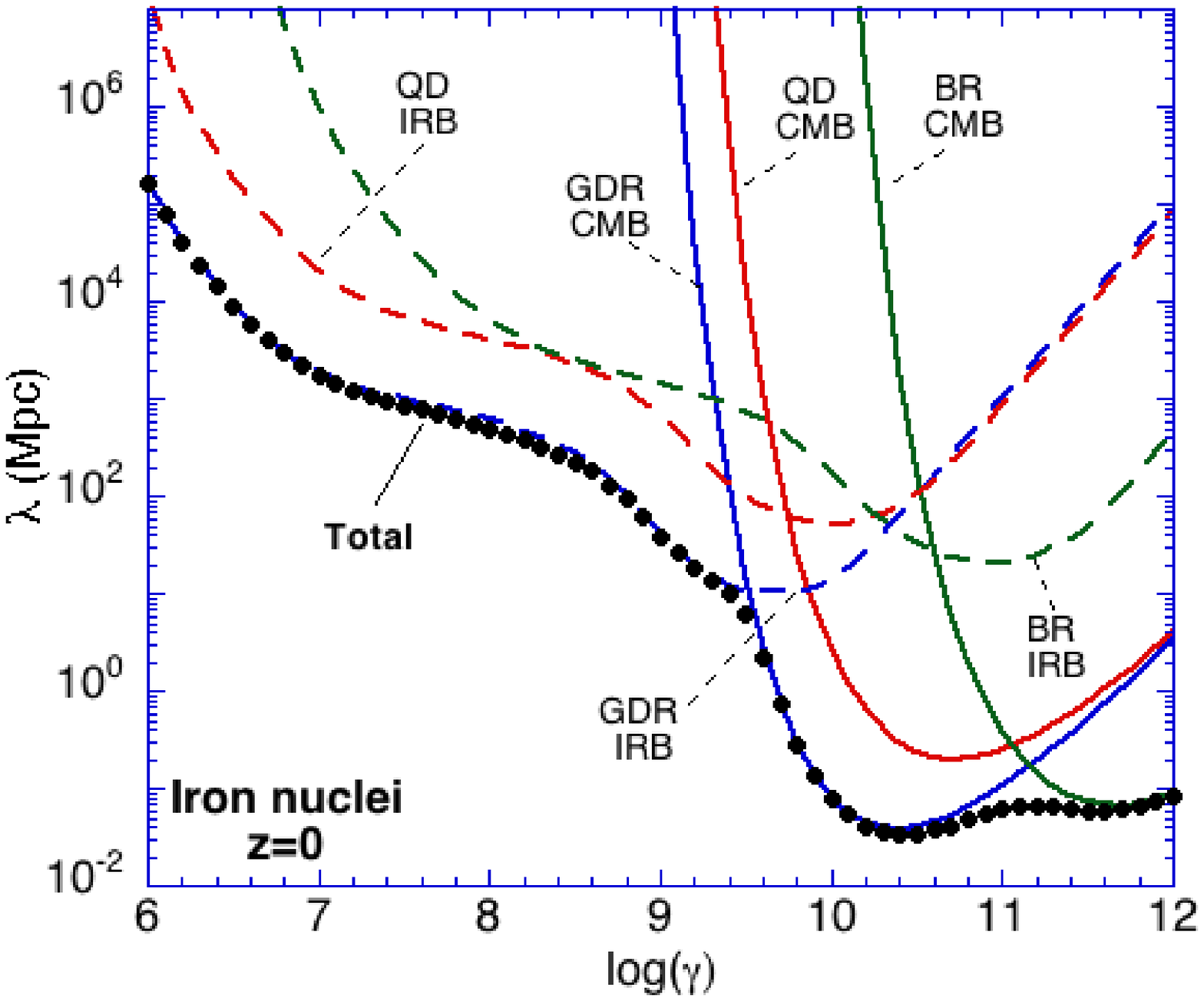}
\hfill\includegraphics[width=0.3\textwidth]{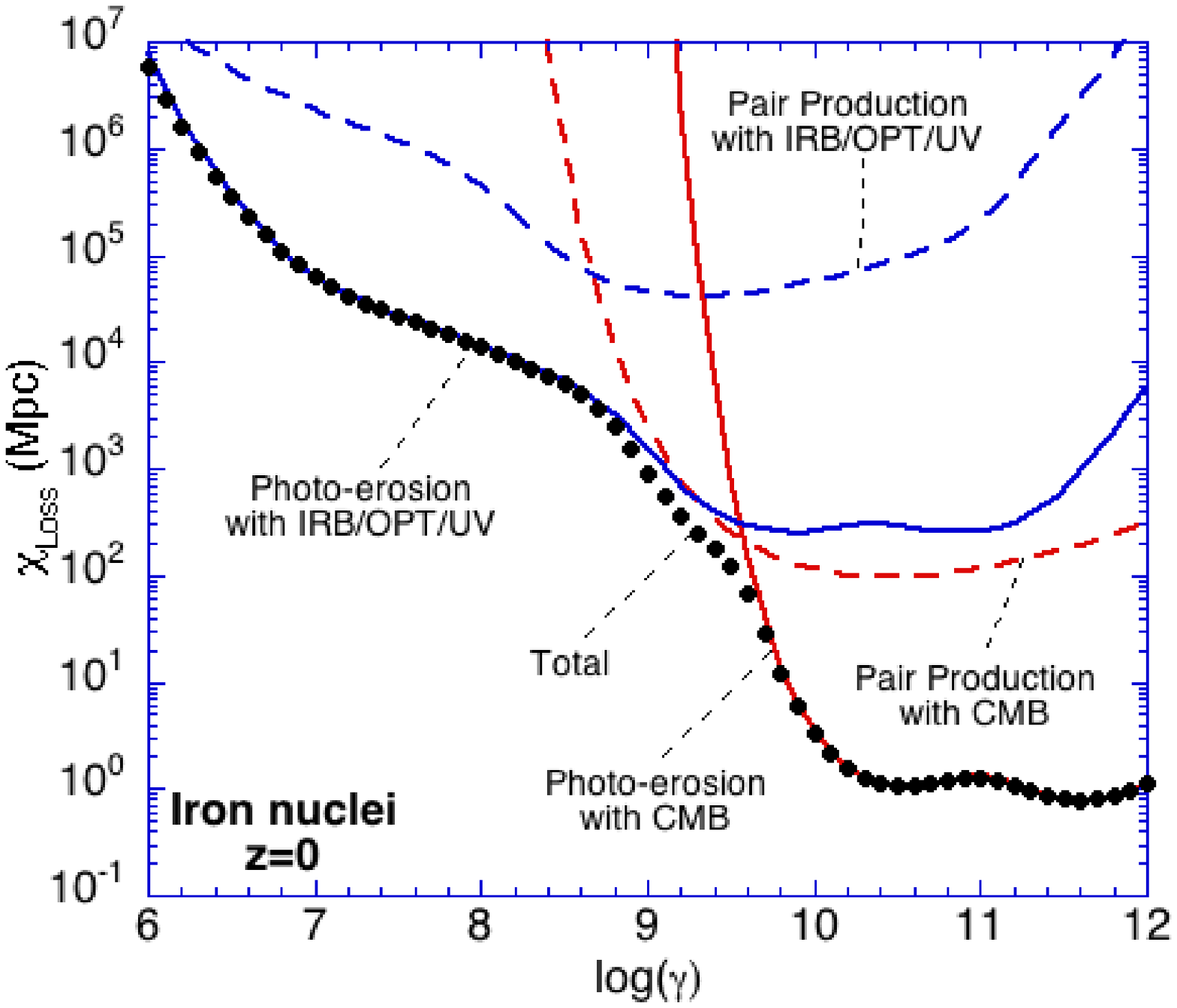}
\hfill\includegraphics[width=0.3\textwidth]{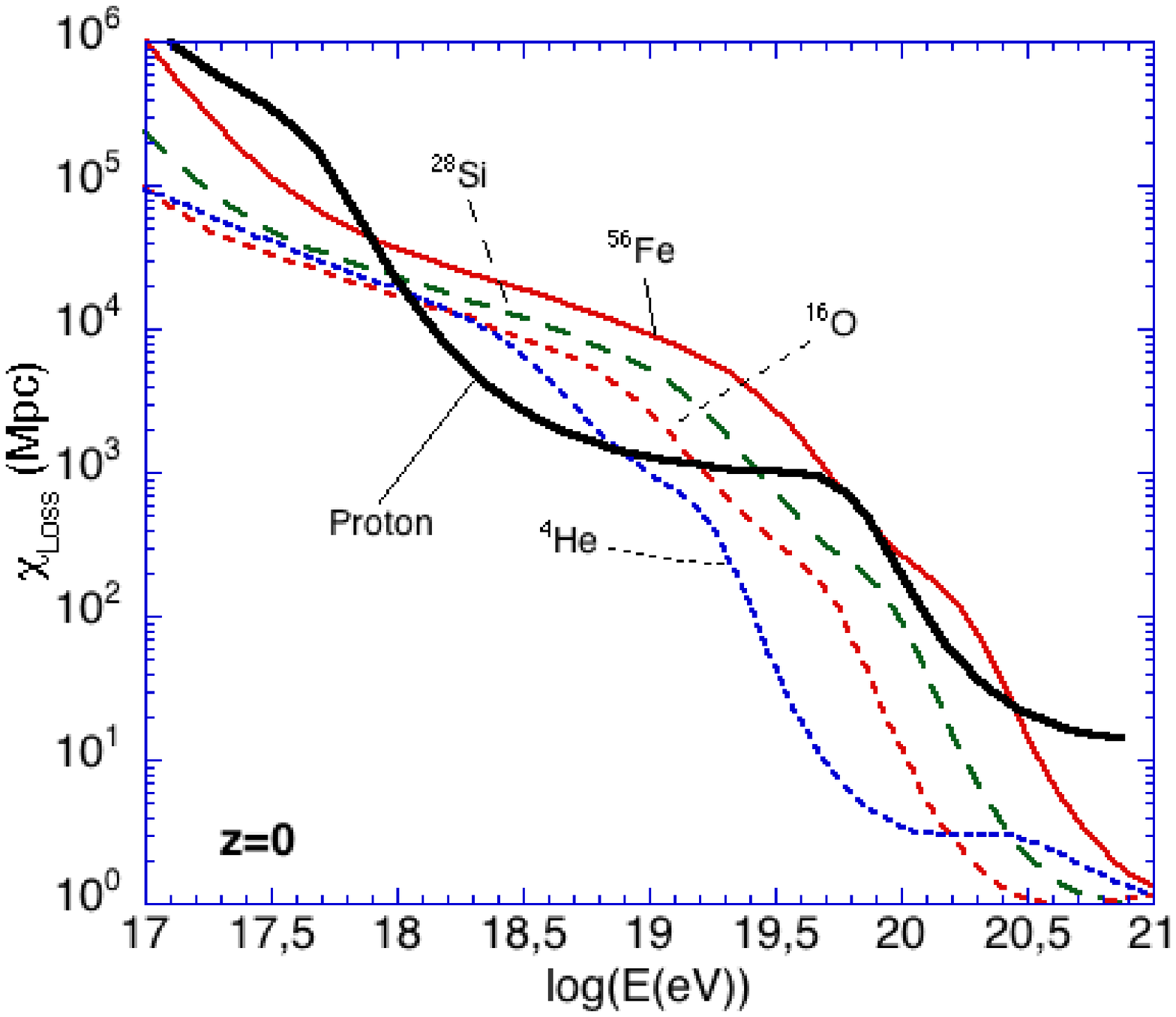}
\hfill~\caption{Left: Evolution of the iron nucleus mean free path as a function of the Lorentz factor for the different photoerosion processes and interactions with the CMB and IR/Opt/UV photons at $z=0$.Center: Evolution of the attenuation length at $z=0$. The contribution of pair production and photoerosion processes off the CMB and IR/Opt/UV photons are separated. Right: Comparison of the energy evolution of the attenuation length of different nuclei at $z=0$.\label{MFP1}}\end{figure*}

The contribution of the different photoerosion processes and the different backgrounds to the total mean free path for iron nuclei are displayed in Fig.~\ref{MFP1}a. The photoerosion is dominated by the GDR process  through most of the Lorentz factor range. The baryonic resonances begin to dominate only above $10^{21.5}$ eV where the effect of the GDR starts to decrease. 
Fig.~\ref{MFP1}b shows the contribution of pair production and photoerosion processes to the total attenuation length of iron nuclei. Photoerosion processes dominate through most of the energy range and the effect of  pair production is small at low redshifts. Although the competition between pair production off the CMB and photoerosion processes with IR/Opt/UV photons depends on the redshift (e.g., at high redshifts pair production increases due to the stronger evolution of the CMB), the propagation of nuclei is mainly dominated by photoerosion processes.  A comparison between the attenuation lengths of different species is displayed in Fig.~\ref{MFP1}c. The figure shows what is
known since \cite{PSB}, that the attenuation length of low mass nuclei are smaller than that of protons and heavy nuclei above $10^{19}$ eV and, as a consequence, light nuclei should not  contribute as significantly at  the high energy end of the spectrum. Furthermore,  iron nuclei have larger or similar attenuation lengths  to protons up to $\sim 3 \times 10^{20}$ eV. However, the energy loss processes are different for protons and nuclei and the sole comparison of attenuation lengths can be misleading and is not straightforward to interpret : most of the energy losses of nuclei result in nucleon ejection, thus, unlike protons, a given nucleus does not remain on ``the same attenuation length curve" during its propagation.

\section{Propagated spectra}
For the following calculation we use a Monte-Carlo code to propagate nuclei from the source to Earth as described in detail in  \cite{Allard05a,Allard2006}. We consider the classical pure proton scenarios and the extragalactic mixed composition models. For our generic mixed nuclei case (see \cite{Allard05a} for more details), we assume that the EGCR source composition matches that of the GCRs observed at lower energies, and that the maximum energy achieved by nuclei of species $i$ in EGCR sources scales with their charge $Z_{i}$, i.e. $E_{\max,i}=Z_{i}  E_{\max}(^{1}H)$, as expected if the acceleration mechanism is controlled by magnetic confinement and limited by particle escape. In the following, we assume a power-law source spectrum with spectral index $\beta$ and set the maximum proton energy to $10^{20.5}$~eV, unless otherwise specified, referring to \cite{Allard2007a} for a discussion on the influence of $E_{max}$. 

We find that the observed UHECR spectra are best fitted with spectral indices between 2.1 and 2.3, which corresponds to a proton dominated composition with significant fractions of He and CNO, and a  lower fraction of heavier nuclei \cite{Allard2006}. However, another important ingredient of EGCR models is the time evolution of the power and/or number density of sources. Indeed, the link between the spectrum of the sources and the observed one (and thus the determination of the ``best fit spectral indices") depends strongly on the assumed redshift evolution of the sources. Here, we consider three different source evolution models. The first one corresponds to no evolution at all -- hereafter referred to as the uniform source distribution model. In the so-called ``SFR model'', we assume that the EGCR injection power is proportional to the star formation rate,  which correspond a to redshift evolution in $(1+z)^3$ for $ z < 1.3$ and a constant injection rate for $1.3 < z < 6$ (with a sharp cutoff at $z = 6$). Finally, we consider a stronger source evolution favoured by the recent infra-red survey of the Spitzer telescope. In this so-called  ``strong evolution model'', we assume a injection rate proportional to $(1+z)^4$ for $ z < 1$ and a constant rate for $1 < z < 6$, followed by a sharp cut-off (see \cite{Allard2006} for more details and references on the sources evolution models).

\begin{figure*}[t]
\centering
\hfill\includegraphics[height=6cm]{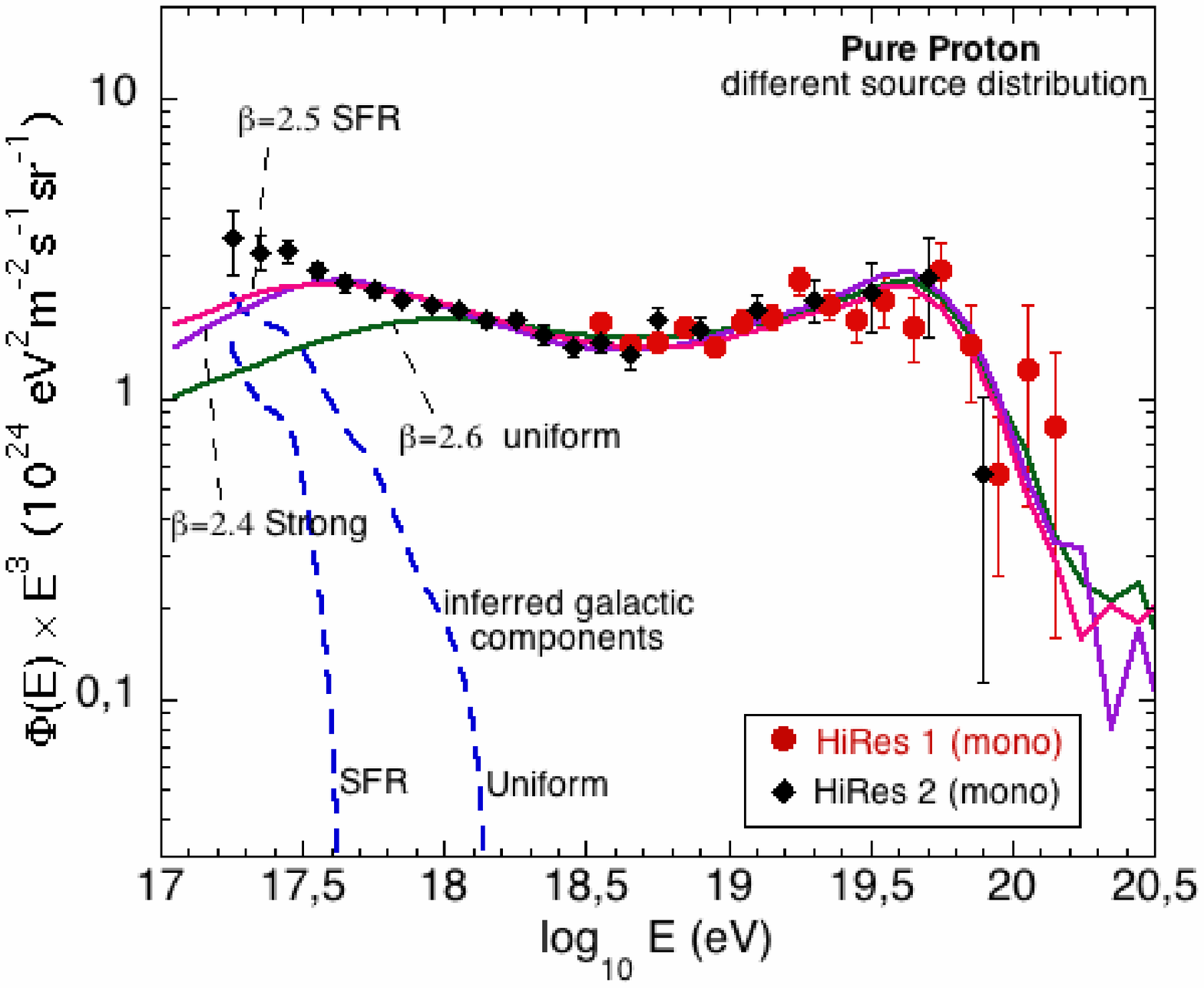}\hfill
\includegraphics[height=6cm]{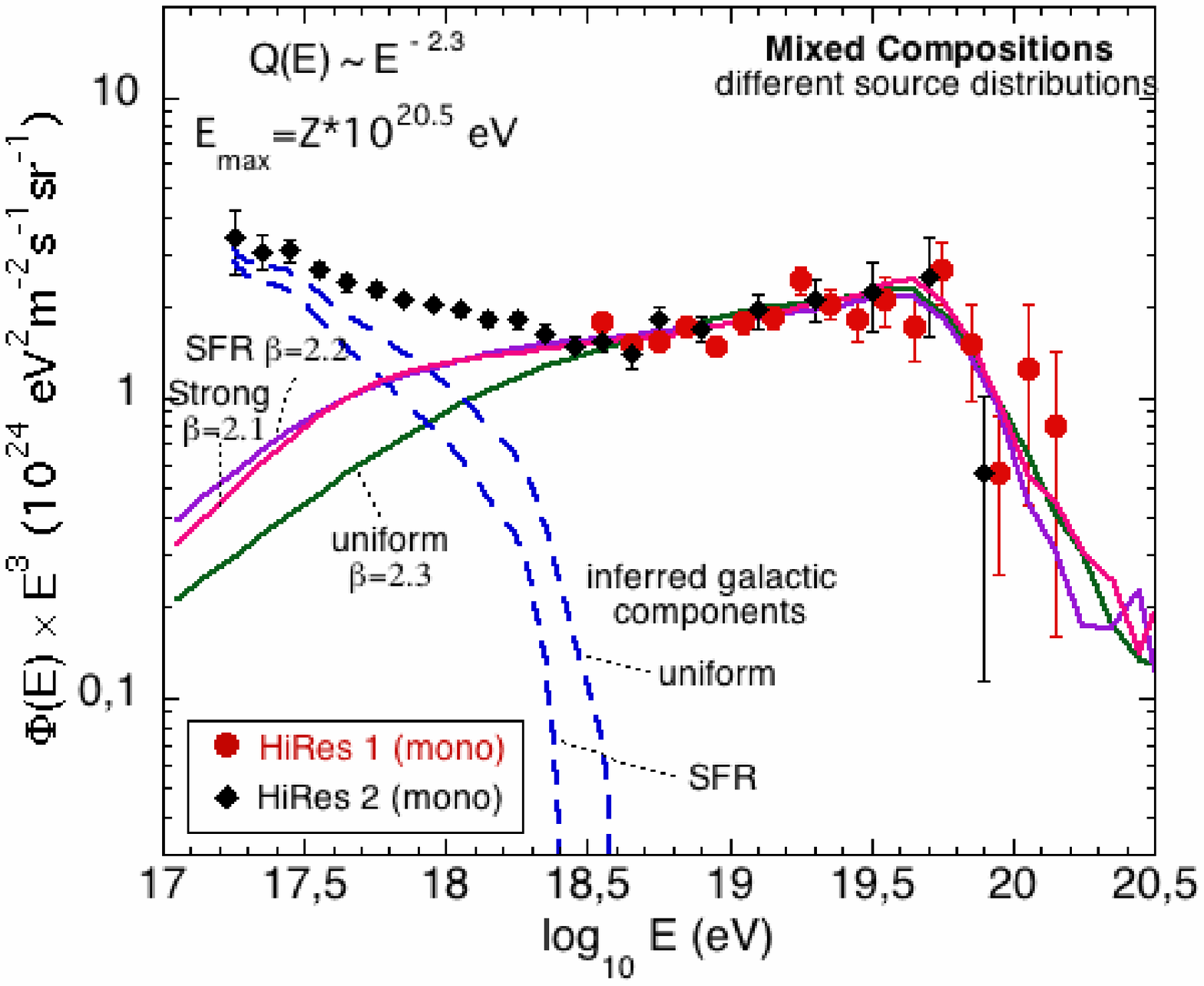}
\hfill~\caption{Propagated spectra, $E^3\Phi(E)$, for pure proton models (left) and mixed composition models (right) compared with HiRes monocular data. Different source evolution models are indicated by the labels. The corresponding galactic components are inferred from the overall spectrum by subtracting the EGCR component, in the case of the uniform and SFR source evolution models.\label{Spectra}}\end{figure*}

In the case of pure proton EGCR sources, the best fit $\beta = 2.6$ if one assumes a uniform distribution of sources (no evolution), while it goes down to 2.5 in the case of an SFR-like evolution, and 2.4 in the strong evolution case (see Fig.~\ref{Spectra}a and \cite{Berezinsky+02,DBO03,DDMS05}).  As shown in previous works \cite{Berezinsky+05,DDMS05}, the propagated proton spectrum and the concave shape known as the ``pair production dip'' (with a minimum around $10^{18.7}$ eV on Fig.~\ref{Spectra}a, for the uniform source model) are only mildly dependent on the source evolution hypothesis.  However, the energy where this e$^{+}$--e$^{-}$ dip begins depends on the relative weight of energy losses related to pair production, which dominate at high energy, and energy losses associated with the universal expansion, which dominate at low energy (see \cite{Berezinsky+02,Berezinsky+04,Berezinsky+05,Berezinsky+05b} for more details). Therefore, the beginning of the dip depends on the redshift evolution of the source density (or power): the transition between the two energy loss processes occurs at a lower energy in the SFR and strong evolution cases (see Fig.~\ref{Spectra}a). In these cases the extragalactic component can account for the whole CR flux down to much lower energies ($\sim 4\,10^{17}$~eV), which correlatively allows/requires the GCR component to cut at relatively low energies, notably lower than the confinement limit of charged nuclei in the Galaxy. Thus, in the pure proton case, the energy $E_{\mathrm{end}}$ at which the GCR/EGCR transition ends (i.e, above which cosmic-rays are purely extragalactic) depends on the source evolution scenario. The highest value of $E_{\mathrm{end}}$ is obtained in the case of a uniform source distribution, around 1--1.5~$10^{18}$~eV. This energy range is significantly lower than in the case of mixed composition scenarios (see below) -- a distinctive feature that can be used to discriminate between the models, using composition analyses.

The propagated EGCR spectra obtained with a mixed source composition are shown in Fig. ~\ref{Spectra}b. The best fit of the high-energy data is obtained in these cases for significantly smaller spectral indices, i.e., harder source spectra: $\beta \simeq 2.3$ in the uniform case, going down to 2.2 for SFR-like source evolution and 2.1 for the strong evolution model.
In all these mixed composition cases, the end of the GCR/EGCR transition roughly coincides with the ankle \cite{Allard05a,Allard2007a}. Above $10^{18.5}$~eV, the predicted spectrum is quite insensitive to the source distribution, as can be seen in Fig.~\ref{Spectra}b. It is also important to note that the mixed composition models do not imply/require any definite value of the highest energy of cosmic rays in the Galactic component, as long as GCRs represent a sufficiently small fraction of the total spectrum around $E_{ankle}$ not to influence the overall spectrum and composition. Therefore, the Galactic component does not necessarily vanish above $E_{\mathrm{ankle}}$, nor is it required that cosmic rays be accelerated above $E_{\mathrm{ankle}}$ at all.
At energies below the ankle, the inferred fraction of GCRs depends on the source evolution model, just as in the pure proton case (see \cite{Allard2007b} for more details). We stress that the energy at which the Galactic and extragalactic components have an equal contribution to the CR flux lies between $\sim 5\,10^{17}$~eV  and $\sim 10^{18}$~eV . Note also that in our mixed composition models the cosmic rays can be dominated by light nuclei at energies below $10^{18}$~eV, which is a major difference with the GCR/EGCR transition scenario studied in \cite{WW}.

\begin{figure*}[t]
\centering
\hfill\includegraphics[height=6cm]{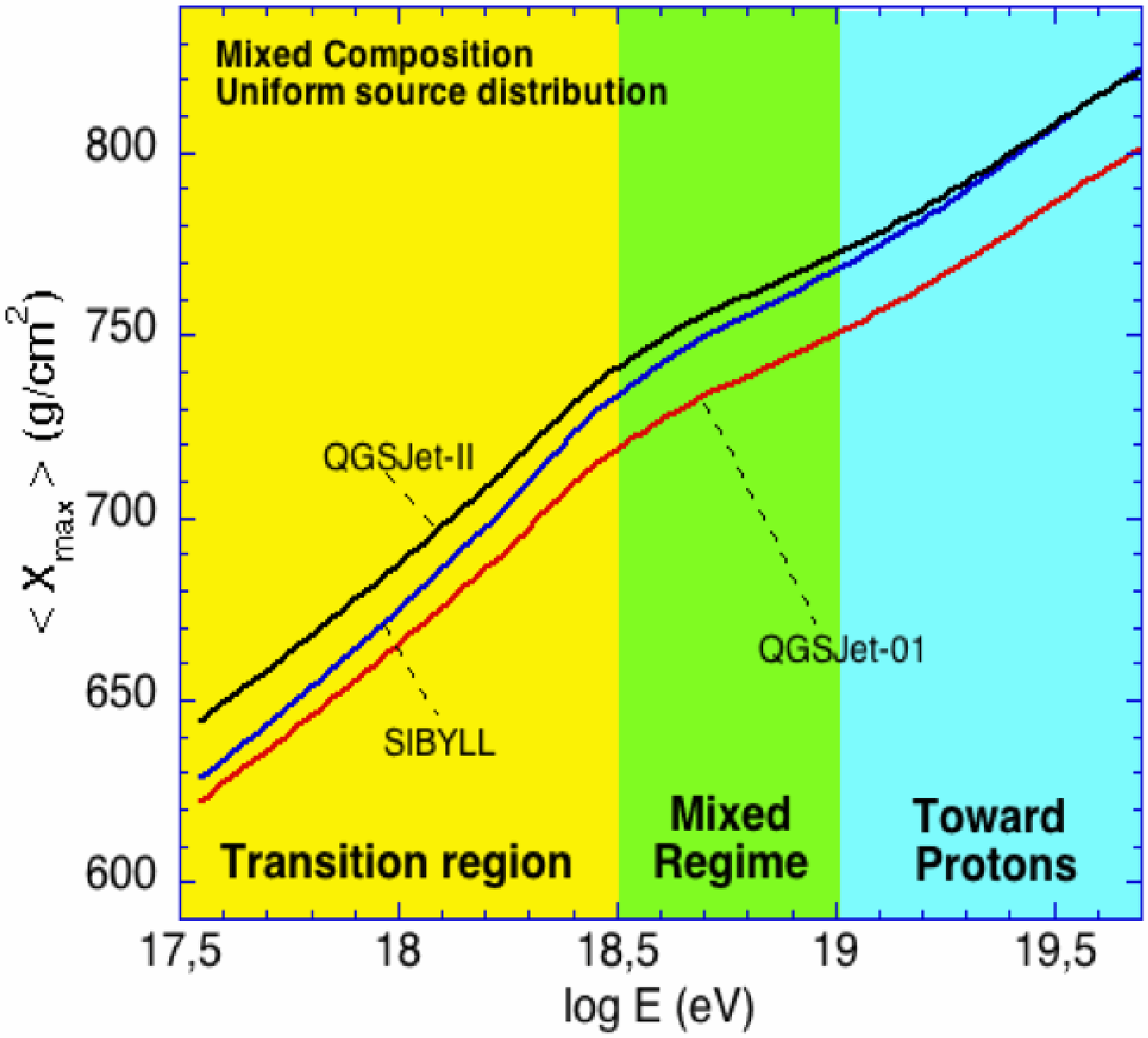}\hfill
\includegraphics[height=6cm]{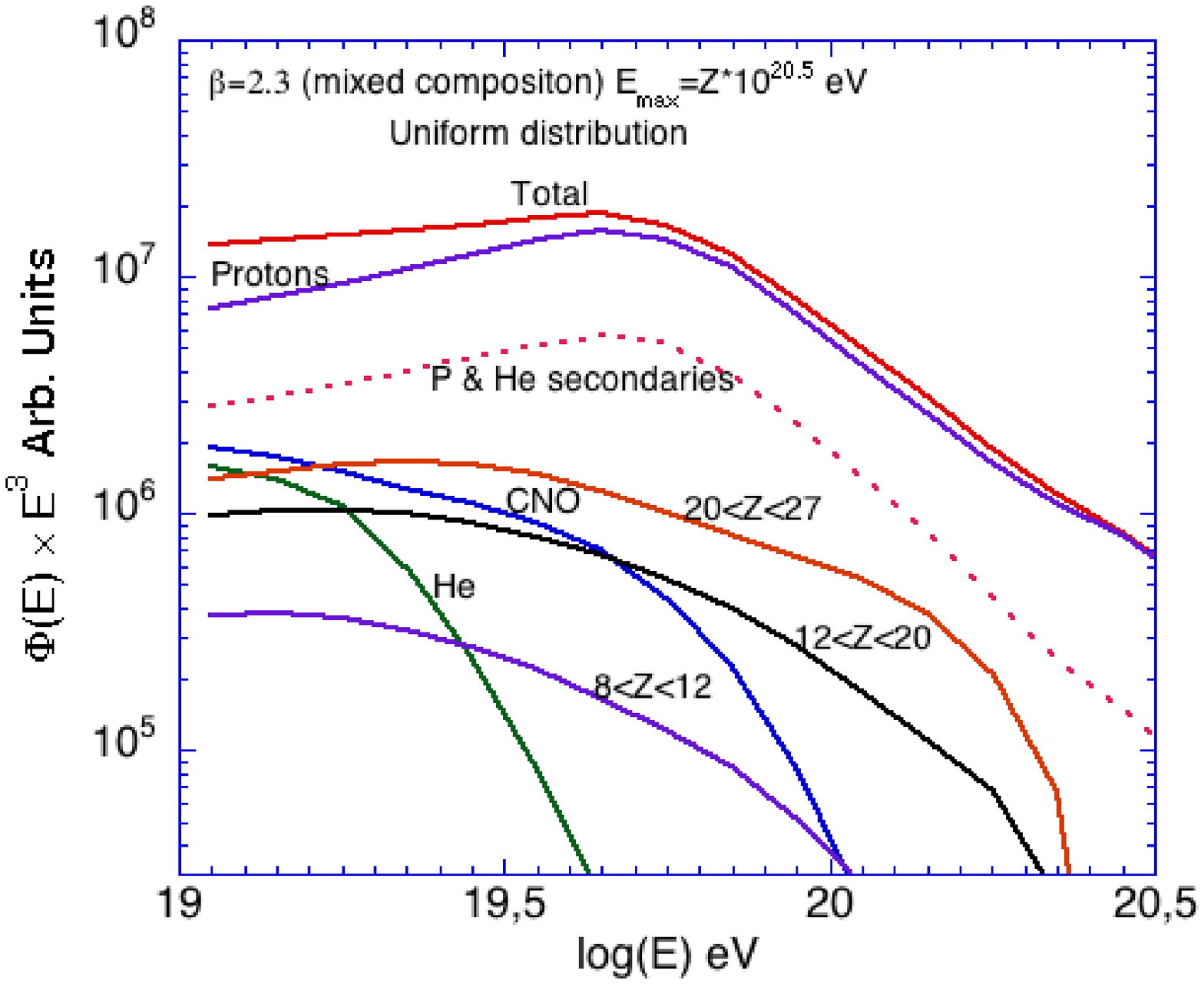}\hfill~
\caption{Left: $\langle X_{\max} \rangle$  evolution for an extragalactic mixed composition and uniform source evolution model for three different hadronic models (see labels). Right: Predicted spectrum for a mixed extragalactic composition above $10^{19}$ eV decomposed in its elemental components. }
\label{Erate2}
\end{figure*}

\section{The shape of $\langle X_{\max} \rangle$(E)}

Having identified the value of $\beta$ that provides the best fit of the data in each scenario, and obtained the corresponding fractions of GCRs and EGCRs at all energies, we can now deduce the evolution of the cosmic-ray composition as a function of energy and predict the values of the associated observables. The \emph{propagated} EGCR composition in each case is a direct output of our computations, and we assume that the Galactic component is essentially made of Fe nuclei above $10^{17.5}$~eV (relaxing this assumption would slightly flatten the $\langle X_{\max} \rangle$ evolution in the transition region). From the relative abundance of all elements at a given energy, we derive the average value of the atmospheric depth (in g/cm$^{2}$) at which the maximum shower development is reached, $\langle X_{\max} \rangle$, using Monte-Carlo shower development simulations \cite{Allard2007a,Allard2007b,Denis2008}.

In the pure proton case, the interpretation of the evolution of $\langle X_{\max} \rangle$ with energy is straightforward (figures are shown in \cite{Allard2007a,Allard2007b}). The transition from Galactic iron nuclei to extragalactic protons being quite narrow (i.e., it occurs over a small energy range, in a decade or even half a decade), the evolution of $\langle X_{\max} \rangle$ with energy is very steep and then gets flatter when the transition is over and the composition does not change anymore, all EGCRs being merely protons. The point where the $\langle X_{\max} \rangle$ evolution can be observed to break simply indicates the energy $E_{\mathrm{end}}$, corresponding to the end of the transition. Characteristically, an early break in the elongation rate at $\sim 4\,10^{17}$~eV is expected in the strong and SFR source evolution models, whereas the break is found around 1--1.5~$10^{18}$~eV for a uniform source distribution. No break is expected at the ankle. Indeed, the ankle is consistently interpreted in pure proton models as the signature of the interactions between EGCR protons and CMB photons producing e$^{+}$e$^{-}$ pairs. Obviously, the resulting ``pair production dip'' would not be visible if the EGCR component did not consist almost exclusively of protons. Quantitatively, nuclei heavier than H cannot contaminate the EGCR component at a higher level than $\sim 15$~\%\cite{WW,Berezinsky+05,Allard05a}.

The case of mixed composition models is illustrated in Fig.~\ref{Erate2}a. The evolution of $\langle X_{\max} \rangle$ is relatively steep in the transition region, below $E_{ankle}$, because the composition evolves rapidly from the dominantly heavy Galactic component to the light extragalactic mixed composition. However, the evolution is significantly slower than in the case of pure proton models, because the transition is wider and the cosmic-ray composition does not turn directly into protons only. As can be seen on Fig.~\ref{Erate2}a, an intermediate stage appears, which may be called the mixed-composition regime, where a break in the evolution of $\langle X_{\max} \rangle$ around $E_{ankle}$ is followed by a flattening up to $\sim10^{19}$~eV, reflecting the fact that the (propagated) EGCR composition does not change much in this energy range. This is because among the different EGCR nuclei, only He nuclei interact strongly with infrared photons at these energies. Between $E_{ankle}$ and $\sim 10^{19}$~eV, the evolution of $\langle X_{\max} \rangle$ is actually compatible with what is expected from a constant composition. Then around $10^{19}$~eV, the relative abundance of nuclei heavier than protons starts to decrease significantly as a result of photo-disintegration processes: the CNO component starts interacting with the infrared background and the CMB photons eventually cause the He component to drop off completely (see Fig.~\ref{Erate2}b). The evolution of $\langle X_{\max} \rangle$ therefore steepens again, accompanying the progressive evolution towards an almost pure proton composition as each type of nuclei reaches its effective (mass dependent) photo-disintegration threshold. Even though slight differences may be expected from one model to the other, the above evolution of $\langle X_{\max} \rangle(E)$ in a three steps process is a characteristic prediction of mixed-composition models, or generically of any type of EGCR sources allowing for the acceleration of a significant fraction of nuclei heavier than He.

\begin{figure*}[t]
\centering
\hfill\includegraphics[height=6cm]{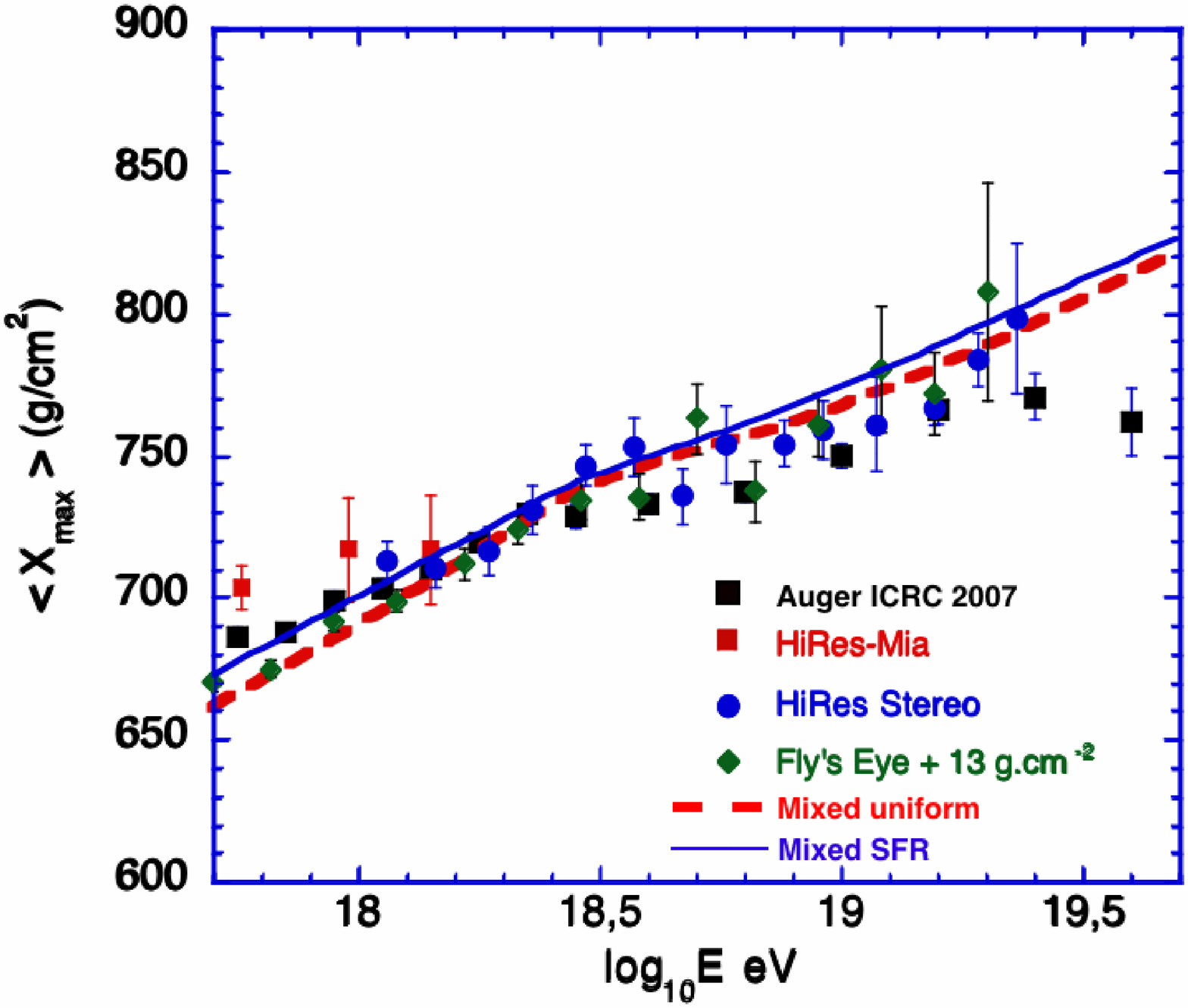}\hfill
\includegraphics[height=6cm]{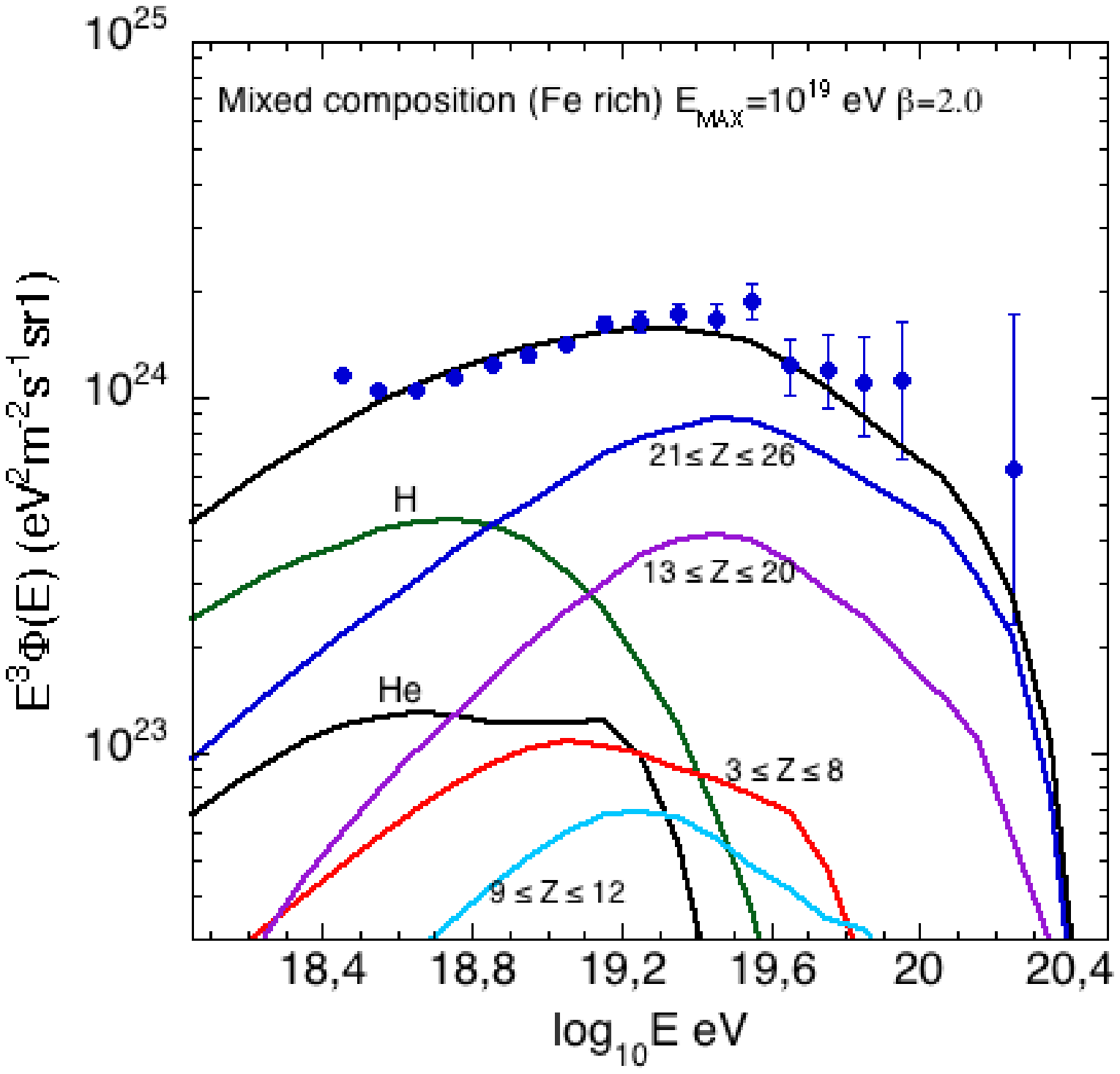}\hfill~
\caption{Left: $\langle X_{\max} \rangle$  evolution for an extragalactic mixed composition above  $10^{19}$ eV, for the SFR and uniform source evolution compared to cosmic-ray data (see legend). Right: Predicted spectrum for a mixed extragalactic composition  ($E_{max}=Z\times10^{19}$ eV) decomposed in its elemental components and compared to Auger data. }
\label{Erate3}
\end{figure*}

\section{Comparison with data and discussion}
The high-energy cosmic ray spectrum can be satisfactorily accounted for within either the pure proton or the mixed composition models (and many other source composition models, see \cite{Allard2008}). However, we have shown that the corresponding phenomenology of the GCR/EGCR transition is very different in each case, which results in distinct shape of $\langle X_{\max} \rangle$ as a function of energy.
The currently available data do not allow one to draw definitive conclusions yet. However, we argued in \cite{Allard2007a} that the predictions of the mixed-composition models appear to be in better agreement with the current data from fluorescence detectors. In particular, a good agreement is found with Fly's Eye results above $10^{17.5}$~eV \cite{Bird+93}. Concerning the slope of the $\langle X_{\max} \rangle$ evolution in the transition region (i.e., below the ankle), mixed-composition models typically predict values for the slope of the  $\langle X_{\max} \rangle$ evolution compatible with what is observed. Furthermore, both the predicted break at the ankle and the steepening above $10^{19}$~eV are compatible with the HiRes Stereo data \cite{HiRescomp}. Fig.~\ref{Erate3}a shows the comparison between the  $\langle X_{\max} \rangle$ evolution for a mixed composition in the SFR evolution case and the data of HiRes Stereo, HiRes-Mia, Fly's Eye (rescaled by 13 $g\,cm^{-2}$, as suggested in \cite{Sok05}) and Auger \cite{Michi}. As can be seen, Fly's Eye and Stereo HiRes data are consistent with the predicted break in the $\langle X_{\max} \rangle$ evolution between 3 and 4~EeV, which also corresponds to the energy of the ankle reported by both experiments. Note that HiRes-Mia results at lower energy are also compatible with pure proton models, as well as with mixed-composition models except at one point around $5\,10^{17}$~eV. The data best agree with the absolute scale of $\langle X_{\max} \rangle$ computed with the QGSJet-II model. However, it is important to note that the results obtained with QGSJet01 show exactly the same features shifted downwards by $\sim$20 $g\,cm^{-2}$, and are still well within the systematic uncertainties of the different experiments. This illustrates once more that the choice of the hadronic model is not critical in the present discussion \cite{Allard2007b}.

When comparing the mixed composition model predictions with the recent data of the Pierre Auger observatory, the agreement appears less good. Auger data are compatible with a break in the $\langle X_{\max} \rangle$ evolution in an energy range close to the ankle (which is, as mentioned before, difficult to handle for a pure proton model) and the overall shape of the evolution below $10^{19}$ eV, although flatter, is compatible with the mixed model predictions. Above $\sim 10^{19}$ eV, however, the composition seems to be getting heavier where the mixed model would predict a lightening. This trend, if confirmed by a larger statistics, would then represent a major incompatibility with our "generic" mixed composition model. Though the reality of this trend is still quite speculative due to the low statistics accumulated at the highest energies, it is interesting to investigate what could cause the composition to get heavier at the highest energies.

In the case of our generic mixed composition, one actually expects that the composition could get heavier above $5\times 10^{19}$ eV if heavy nuclei are accelerated at the highest energies. Indeed, above energy, the proton flux decreases sharply due to photopion interactions with CMB photons. Between $5\times 10^{19}$ eV and $\sim2.5\times 10^{20}$ eV, heavy nuclei (Fe) only interact with the less dense far-IR photons and the decrease of their flux is then slower (see Fig.~\ref{Erate2}b) than for protons, implying an increase of their relative abundance. This trend ends when the heavy component disappears due to interaction with CMB photons (see \cite{Allard2007b} for more details). However, the energy range where this feature is expected does not seem to be compatible with Auger data. Invoking a heavy dominated or possibly pure iron composition at the sources would presumably not solve the problem either. The composition analyses at lower energy seem to favor a transition from heavy galactic to light extragalactic cosmic-rays which does not look compatible with the expectations of an extragalactic heavy source composition (see discussion of the pure iron sources in \cite{Allard2007a}). Let us note that the latter point is somehow alleviated but remains true if one assumes a strong evolution of source luminosity with redshift for which spectral indices of 2.0-2.1 would be required to fit the observed spectra. 

In this context, a possible explanation for this trend would be to infer that most of the sources are not able to accelerate protons up to the highest energies and that only heavy nuclei are accelerated above $10^{20}$ eV (in the case of a maximum energy scaling with the charge of the nucleus, this kind of scenario is proposed  for instance in \cite{Inoue2007}). As we discussed in \cite{Allard2007a}, low $E_{max}$ proton solutions do not work very well with our usual mixed composition hypothesis.  Some tuning of the composition is then necessary for this type of scenario to be compatible with  the data. However, good fits of the data can be obtained with a moderate increase of the overall abundance of heavy elements at the source, by a factor of three or so with respect to low-energy Galactic cosmic rays. This is illustrated in Fig.~\ref{Erate3}b where expected spectra are displayed and compared with data assuming a mixed composition, $\beta=2.0$ , $E_{max}=Z\times10^{19}$ eV and $\sim30\%$ of Fe nuclei at the sources. One can see that the agreement with data is reasonable and that the composition, proton dominated at low energy, becomes gradually heavier and very dominated by iron above $5\,10^{19}$ eV. Such a scenario would have implications on the cosmogenic neutrino flux that should be extremely low above $10^{17}$ eV. Neutrinos and photons fluxes from interactions during the acceleration at the sources would as well be presumably very low (see discussion in \cite{AllardProtheroe2009}). To conclude, let us note that, at the current level of statistics, it seems difficult to argue that the recent anisotropy claim by the Pierre Auger collaboration \cite{Correlation} disfavors heavy compositions at the highest energie (see for instance \cite{WW2}). However, a significant small scale clustering, if observed in the future, would challenge this type of scenario.
 
Although the available cosmic-ray data already allow to put encouraging constrains on the different models we presented, a clear picture of the GCR to EGCR transition, the composition and the origin of the UHECR is still difficult to draw.  Future data of the Pierre Auger Observatory, its low energy extension as well as the future very large aperture projects such as Auger North \cite{Dave07} and JEM-EUSO \cite{EUSO} should allow a clearer view on these questions in the next few years.

\section*{Acknowledgements}
D. A. wishes to thank A. Olinto, E. Parizot, N. G. Busca, N. Globus, M. Ave, T. Yamamoto, M. Malkan, F. W. Stecker, E. Khan, S. Goriely, G. Decerprit for their collaboration on the work described in this paper.

\end{document}